\documentclass{aastex}
\usepackage{spr-astr-addons}
\usepackage{url}
\usepackage{multirow}

\begin{document}

\title{A complex asteroseismic study of\\the hybrid B-type pulsator $\nu$ Eridani}
\author{J. Daszy\'nska-Daszkiewicz, P. Walczak}
\affil{Instytut Astronomiczny, Uniwersytet Wroc{\l}awski, ul. Kopernika 11, 51-622 Wroc{\l}aw, POLAND}

\shorttitle{A complex asteroseismic study of $\nu$ Eridani}
\shortauthors{Daszy\'nska-Daszkiewicz \& Walczak}

\begin{abstract}
We present results of comprehensive seismic modelling of the B-type main-sequence pulsator $\nu$ Eridani,
which consists in parallel fitting of the pulsational frequencies and corresponding values
of the complex, nonadiabatic parameter $f$, defined by the bolometric flux perturbation.
This kind of studies, which we call {\it complex asteroseismology}, provides
a unique test of stellar opacities. Our results indicate a preference for the OPAL data.
\end{abstract}


\section{Introduction}
$\nu$ Eridani (HD 29248) is a well known $\beta$ Cep/SPB type pulsator with the B2III spectral type
and brightness of $m_V$=3.92 mag. The star is a slow rotator with the velocity of about 6 km/s
derived from a rotational splitting of pulsational frequencies.
In the last years, this variable has attracted a lot of interest thanks to dedicated photometric
(Handler et al. 2004, Jerzykiewicz et al. 2005) and spectroscopic (Aert et al. 2004) multisite campaigns.
The frequency analysis revealed 14 independent peaks in photometry (Handler 2004, Jerzykiewicz et al. 2005)
and 9 of them were identified also in spectroscopy (Aerts et al. 2004, de Ridder et al. 2004).
Subsequently, several papers were devoted to seismic analysis of $\nu$ Eri.
Pamyatnykh, Handler \& Dziembowski (2004) presented the first seismic modelling
of this star with a special emphasis of excitation problem for high-order g mode.
Ausseloos et al. (2004) managed to fit four frequencies but for models with very low effective
temperature (far outside the $3\sigma$ error) or unacceptable chemical composition.
Daszy\'nska-Daszkiewicz et al. (2005) undertook the first attempt to compare the empirical
and theoretical values of $f$ for the radial mode, but neither with OP nor with OPAL opacity tables
an agreement was achieved. More recently, Dziembowski \& Pamyatnykh (2008)
redid seismic modelling with the new solar chemical composition as determined by Asplund et al.
(2004, hereafter A04) but again no satisfactory interpretation of the whole $\nu$ Eri spectrum was obtained.
Effects of differential rotation in the analysis of rotational splitting of three $\ell=1$ modes
and their asymmetries were included by Suarez et al. (2009). A few years ago, the asymmetry of one $\ell=1$ triplet
was suggested to be caused by a strong magnetic field (Dziembowski\& Jerzykiewicz 2003), but its presence was
not confirmed (Schnerr et al. 2006).

A position of $\nu$ Eri in the HR diagram is shown in the right panel of Fig.\,1.
The observational values of the effective temperature and luminosity
were taken from Pamyatnykh, Handler \& Dziembowski (2004).
The evolutionary tracks, shown from ZAMS to TAMS, were computed with the Warsaw-New Jersey evolutionary
code adopting the OP opacities and the A04 solar mixture.
We assumed the rotational velocity of 10 km/s and hydrogen abundance of $X=0.7$ at ZAMS.
The effect of the heavy elements abundance, $Z$, and the overshooting parameter, $\alpha_{\rm{ov}}$,
on the evolutionary tracks is also presented.
Other features of this figure will be discussed later on.

Here, we present mode identification for all detected pulsational frequencies
as well as one more attempt towards seismic modelling of this B-type hybrid pulsator.
In our studies we aim at simultaneous fitting of the pulsational frequencies and
corresponding values of the complex, nonadiabatic parameter $f$, taking into account
instability conditions.
All pulsation computations were done using the linear nonadiabatic code of Dziembowski (1977).
The $f$-parameter is the complex photospheric amplitude of the radiative flux perturbation and,
in the case of B-type pulsators, its value is very sensitive to metal abundance and opacities
(Daszy\'nska-Daszkiewicz et al. 2005). Such extended seismic study has been recently done for
the $\beta$ Cep star $\theta$ Ophiuchi (Daszy\'nska-Daszkiewicz \& Walczak 2009).
In these proceedings we give only a very brief outline of our results.
The full analysis will be published elsewhere.
\begin{table*}[t]
\small
\begin{center}
\caption{The most probable identification of $\ell$ for the pulsational frequencies
of $\nu$ Eri from two methods, denoted as $\ell_1$ and $\ell_2$ (see text for explanation).
In the next four columns, values of amplitudes and phases
of the $y$ photometric and radial velocity variation are given.
The last two columns contain the real and imaginary parts of the empirical values of the $f$-parameter
for the pulsational frequencies detected in both photometry and spectroscopy.
Stellar parameters of the OPAL seismic model marked with the star symbol in Fig.\,1 were assumed:
$M=9.48~M_{\odot},~ \log T_{\rm eff}=4.346,~\log L/L_{\odot}=3.924$.}
\begin{tabular}{|c|c|c|c|c|c|c|c|c|}
\hline

frequency  &\multirow{2}{*}{$\ell_{1}$} &\multirow{2}{*}{$\ell_{2}$}&$A_{\rm{y}}$&$\phi_{\rm{y}}$&$V_{\rm{rad}}$&$\phi_{\rm{Vrad}}$& \multirow{2}{*}{$f_{\rm{R}}$}&\multirow{2}{*}{$f_{\rm{I}}$}\\

[c/d]    & &  &[mmag]      &[rad]          &[km/s]        & [rad]            &    &
\\  \hline
$\nu_{1}$=5.76328 &$\ell$=0 &$\ell$=0  & 36.88(12)&1.032(3)  &22.02(12)&2.782(5)&-9.17(30)&0.76(31)	 \\
$\nu_{2}$=5.65388 &$\ell$=1 &$\ell$=1  & 25.41(13)&0.250(5)  &9.05(12)&1.841(13)&-9.03(25)&0.15(25)	 \\
$\nu_{3}$=5.62002 &$\ell$=1 &$\ell$=1  & 22.70(13)&2.287(6)  &8.27(11)&3.859(14)&-8.82(30)&0.02(31)	 \\
$\nu_{4}$=5.63725 &$\ell$=1 &$\ell$=1  & 21.21(14)&3.916(6)  &7.70(11)&5.543(14)&-8.90(16)&0.51(16)	 \\
$\nu_{5}$=7.8982  &$\ell$=1 &$\ell$=1  & ~2.75(13)&3.832(47)  &0.99(11)&5.804(111)&-11.04(78)&4.43(78)	 \\
$\nu_{6}$=6.24385 &$\ell$=1 &$\ell$=1  & ~2.40(13)&6.047(52) &0.92(10)&1.131(114)&-8.88(1.20)&-1.25(1.20)	 \\
$\nu_{7}$=6.26292 &$\ell$=1 &$\ell$=1,3& ~1.59(13)&2.162(80)  &0.84(11)&3.928(128)&-6.92(38)&1.26(38)	 \\
$\nu_{8}$=7.2009  &$\ell$=2 &   -      & ~1.06(12)&4.112(115)&-&-&-&-\\
$\nu_{9}$=7.91383 &$\ell$=1	&   -      & ~0.89(13)&4.252(152)&-&-	&-&-\\
$\nu_{10}$=7.92992&$\ell$=1	&   -      & ~0.40(13)&6.112(321)&-&-	&-&-\\
$\nu_{11}$=6.73223&$\ell$=?	&   -      & ~0.70(12)&3.240(177)&-&-	&-&-\\
$\nu_{12}$=6.2236 &$\ell$=1 &$\ell$=1,2,3&~0.76(13)&0.712(165)&0.43(11)&3.126(247)&-3.95(73)&4.43(74)	 \\
$\nu_{A}$=0.43279 &$\ell$=1	& -        & ~3.15(13)&0.344(37) &-&-	&-&-\\
$\nu_{B}$=0.6144 &$\ell$=2,4&$\ell$=2,5& ~1.41(12)&5.578(85) &0.71(10)&2.398(141)&9.35(4.11)&16.49(4.11)	 \\
\hline
\end{tabular}
\label{tab_mse}
\end{center}
\end{table*}

\section{Mode identification}
Although a few years have elapsed since the second photometric campaign of $\nu$ Eri
(Jerzykiewicz et al. 2005), mode identification employing photometric observables
for all 14 pulsational frequencies has never been undertaken.
The $uvy$ Str\"omgren photometry and radial velocity measurements allowed us to apply
two approaches to identify the mode degree, $\ell$.
In the first case, we compare observational values of the amplitude ratios and phase differences
between available passpands with their theoretical counterparts and rely on theoretical
values of the $f$-parameter, which results from linear
nonadiabatic computations of stellar oscillations (Cugier, Dziembowski \& Pamyatnykh, 1994).
In the second method, one makes use of amplitudes and phases themselves
and the value of $f$ is determined simultaneously with $\ell$ in the Least Square process
(Daszy\'nska-Daszkiewicz et al. 2003, 2005).
In the case of B-type pulsators, the second method demands adding the radial velocity measurements
to get an unambiguous identification of the degree, $\ell$.
In both methods we adopt the Kurucz models of stellar atmospheres.

In Tab.\,1 we give identification of $\ell$ for all pulsational frequencies
of $\nu$ Eri from two approaches; $\ell_1$ and $\ell_2$ correspond to determination
with the theoretical and empirical values of $f$, respectively.
Next columns contain values of the amplitudes and phases in the Str\"omgren $y$ passband
and of radial velocity variation.
As we have mentioned above, from the second method we derive also the complex parameter $f$,
whose exemplary values are given in the last two columns of Tab.\,1.
The empirical values of $f$ can be compared with the results of the nonadiabatic pulsation
computations. In the next section, a requirement of getting an agreement for the $f$-parameter
will be added to seismic modelling.
\begin{figure*}[ht]
\begin{center}
\includegraphics[width=\textwidth,clip]{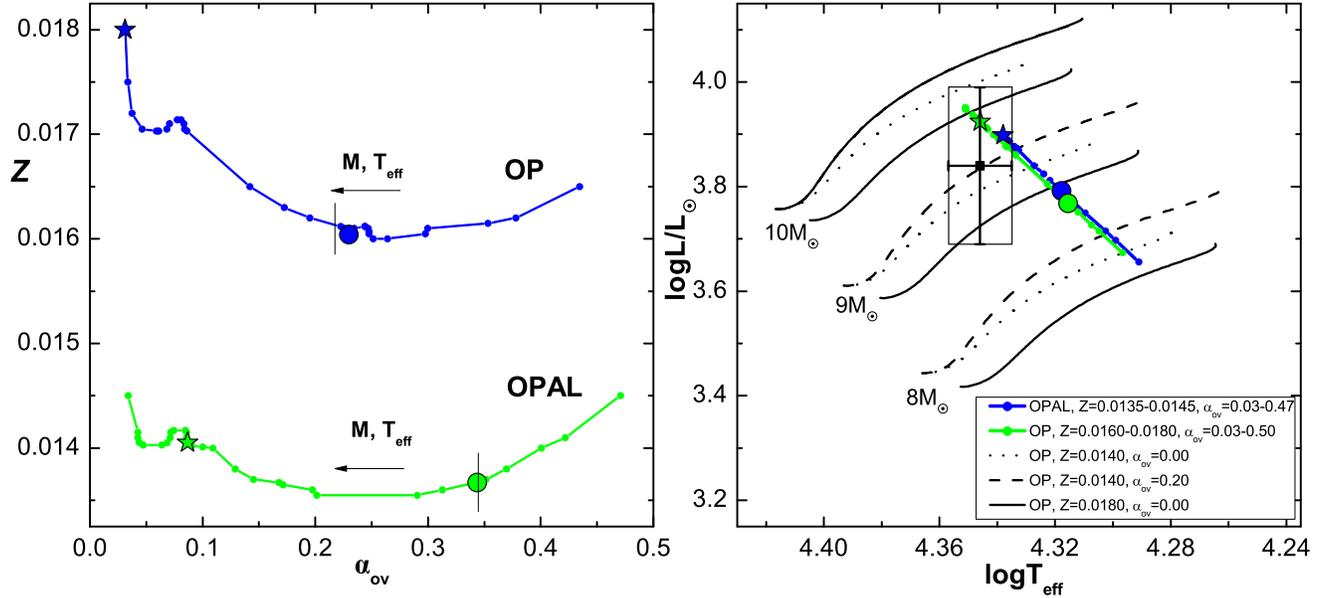}
\caption{On the left: the OP and OPAL seismic models which fit the three centroid frequencies of $\nu$ Eri:
$\nu_1, \nu_4$ and $\nu_6$, on the $Z~vs.~\alpha_{\rm{ov}}$ plane. Models to the left
of a vertical dash have all three modes unstable. The big dot symbols mark models
which fit also a centroid frequency of the fourth triplet, $\nu_{9}$.
Models marked with the star symbols are discussed in Sect.\,3.
On the right: a position of the OP and OPAL seismic models in the HR diagram and the observational
error box of $\nu$ Eri. Evolutionary tracks are described in Introduction.}
\label{}
\end{center}
\end{figure*}
\begin{figure*}[ht]
\begin{center}
\includegraphics[width=\textwidth,clip]{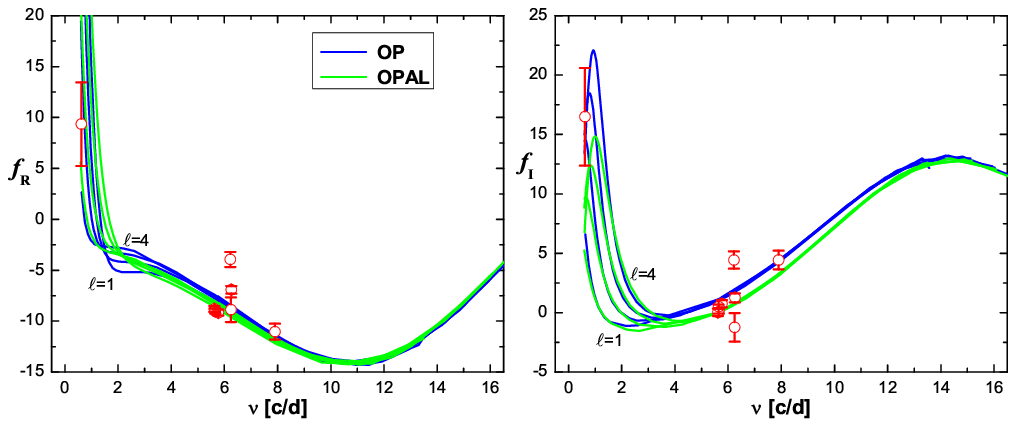}
\caption{Comparison of the empirical and theoretical values of $f$ in the whole range
of frequencies observed in both photometric and spectroscopic variation of $\nu$ Eri.
The real and imaginary parts of $f$ are plotted in the left and right panel, respectively.
Theoretical values of $f$ correspond to the OP and OPAL seismic models marked with
star symbols in Fig.\,1. Modes with the degree, $\ell$, from 0 to 4 were considered.}
             \label{}
             \end{center}
        \end{figure*}

\section{Seismic models of $\nu$ Eridani}
In the left panel of Fig.\,1, we put seismic models of $\nu$ Eridani on the $Z~vs.~\alpha_{\rm ov}$ plane.
All these models reproduce three centroid frequencies: $\nu_{1}(\ell=0,{\rm p}_{1})$,
$\nu_{4}(\ell=1,{\rm g}_{1})$ and $\nu_{6}(\ell=1,{\rm p}_{1})$, and were found with the OP and OPAL tables
assuming the A04 solar composition. Models to the left of the vertical thick dash have all these three modes unstable.
Moreover, in this family of seismic models there are ones which fit also the fourth centroid frequency,
$\nu_{9}(\ell=1,{\rm p}_{2})$ (big dots in Fig.\,1), but with a worse accuracy and this mode is stable.
These models are beyond the observational error box but still within the $3\sigma$ error of the effective temperature.
A similar result was obtained by Ausseloos et al. (2004) but for cooler temperatures or rather unrealistic
chemical composition.

The results obtained with the OP and OPAL opacity tables are similar but
the OPAL seismic models require smaller metallicity, have larger masses and
effective temperatures which fit better observational values of the effective temperature, $T_{\rm{eff}}$,
and luminosity, $L$ (see the right panel of Fig.\,1).

For a comparison of the empirical and theoretical values of the $f$-parameter, we chose two seismic models
within the observational error box computed with the OP and OPAL tables (star symbols in Fig.\,1).
In Fig.\,2, we show this comparison for nine frequencies found in both photometry and spectroscopy.
As can be seen, there is an agreement for the most pulsational frequencies
including the SPB-type mode, $\nu_B$=0.6144 c/d. The discrepancy for the remaining frequencies
is related to a low accuracy in amplitudes and phases.
In Fig.\,3, we present a more detailed comparison for the radial mode, $\nu_1$,
using all OP and OPAL seismic models. As we can see, an agreement can be achieved only
with the OPAL data and for cooler models.

\section{Conclusions}
We presented mode identification for all 14 pulsational frequencies detected in the light
variation of $\nu$ Eri. For nine frequencies, visible also in the radial velocity variation,
we were able to apply the method of simultaneous determination of the mode degree, $\ell$,
and the complex, nonadiabatic parameter $f$.
In the next step, we looked for stellar models which fit three centroid frequencies:
$\nu_{1}$, $\nu_{4}$ and $\nu_{6}$.
Using opacities from both the OP and OPAL tables and the A04 chemical mixture,
we found a family of seismic models with different masses, temperatures, metallicities
and core overshooting parameters. Seismic models computed with the OPAL opacities fit
better observational values of $T_{\rm{eff}}$ and $L$.
We managed to find also models which fit the fourth centroid frequency, $\nu_{9}$.
The accuracy of this fitting is worse comparing to the first three frequencies
but these models are within the $3\sigma$ error of $\log T_{\rm eff}$.

Then, we compared the empirical and theoretical values of the $f$-parameter for nine frequencies
which appeared in both photometry and spectroscopy. The overall agreement is very encouraging.
In particular, the obtained concordance for high-order g mode opens a new gate in seismic studies
of the hybrid pulsators of the $\beta$ Cep/SPB type.
Moreover, a detailed comparison of the values of $f$ for the radial fundamental mode showed
that the OPAL tables are clearly favoured. The same result was obtained by Daszy\'nska-Daszkiewicz \& Walczak (2009)
from the analysis of the $\beta$ Cep star $\theta$ Ophiuchi.
\begin{figure}[ht]
\centering
\includegraphics[width=84mm,clip]{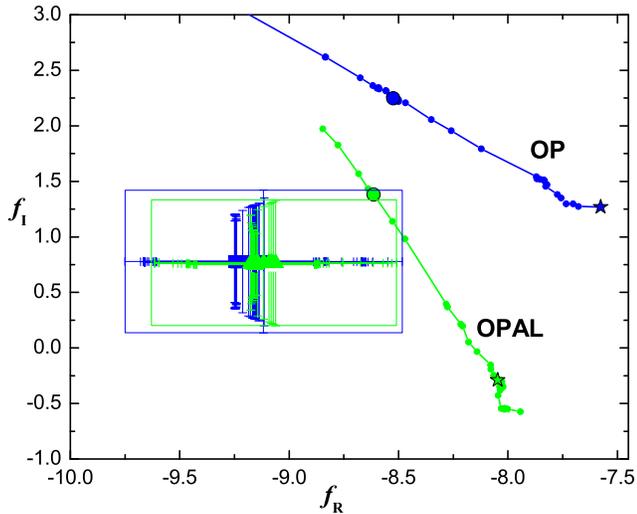}
\caption{Comparison of the empirical values of $f$ for the radial fundamental mode,
$\nu_1$, with theoretical ones corresponding to the OP and OPAL seismic models of $\nu$ Eri.}
\label{aaaaaa}
\end{figure}

\acknowledgments
We thank Mike Jerzykiewicz for the $\nu$ Eri light curves in the Str\"omgren passbands.
The work was supported by the the European Helio- and Asteroseismology
Network (HELAS), No. 026138.

\makeatletter
\let\clear@thebibliography@page=\relax
\makeatother

{}


\begin{thebibliography}{}

\bibitem[\protect\citeauthoryear{{Aerts et al.}}{2004}]{}
  Aerts, ~C., De ~Cat, ~P., Handler, ~G.  et al.:  \mnras\ \textbf{347}, 221 (2004)

\bibitem[\protect\citeauthoryear{{Asplund et al}}{2004}]{}
 Asplund, ~M., Grevesse, ~N., Sauval, ~A.J., Allendre Prieto. ~C., Kiselman, D.: \aap\ \textbf{417}, 751 (2004)

\bibitem[\protect\citeauthoryear{{ Ausseloos et al.}}{2004}]{}
Ausseloos, ~M., Scuflaire, ~R., Thoul, ~A., Aerts, ~C.: \mnras\ \textbf{355}, 352 (2004)

\bibitem[\protect\citeauthoryear{ Cugier et al.}{1994}]{cug94}
Cugier, ~H., Dziembowski, ~W.A., Pamyatnykh, ~A.A.: \aap\ \textbf{291}, 143 (1994)

\bibitem[\protect\citeauthoryear{{Daszy\'nska-Daszkiewicz et al.}}{2003}]{das03}
Daszy\'nska-Daszkiewicz, ~J., Dziembowski, ~W. A., Pamyatnykh,  ~A. A.:  \aap\ \textbf{407}, 999  (2003)

\bibitem[\protect\citeauthoryear{{Daszy\'nska-Daszkiewicz et al}}{2005}]{}
Daszy\'nska-Daszkiewicz, ~J., Dziembowski, ~W. A., Pamyatnykh,  ~A. A.:  \aap\ \textbf{441}, 641  (2005)

\bibitem[\protect\citeauthoryear{{Daszy\'nska-Daszkiewicz et al}}{2009}]{}
Daszy\'nska-Daszkiewicz, ~J., Walczak, ~P.: \mnras\ \textbf{398}, 1961 (2009)

\bibitem[\protect\citeauthoryear{{De Ridder et al}}{2004}]{}
De Ridder, ~J., Telting, ~J. H., Balona, ~L. A.,  et al.: \mnras\ \textbf{351}, 324 (2004)

\bibitem[\protect\citeauthoryear{{Dziembowski}}{1977}]{}
Dziembowski W. A., 1977, \actaa, 27, 203

\bibitem[\protect\citeauthoryear{{Dziembowski et al}}{2008}]{}
Dziembowski, ~W. A., Pamyatnykh,  ~A. A.: \mnras\ \textbf{385}, 2061 (2008)

\bibitem[\protect\citeauthoryear{{Dziembowski et al}}{2003}]{}
Dziembowski, W. A., Jerzykiewicz, M. 2003, in International Conference on
magnetic fields in O, B and A stars, ed. L. A. Balona, H. F. Henrichs,
R. Medupe, ASP Conf. Ser., \textbf{305}, 319

\bibitem[\protect\citeauthoryear{{Handler et al}}{2004}]{}
Handler, ~G., Shobbrook, ~R. R., Jerzykiewicz, ~M., et al.: \mnras \textbf{347}, 454 (2004)

\bibitem[\protect\citeauthoryear{{Jerzykiewicz et al}}{2005}]{}
Jerzykiewicz, ~M., Handler,~ G., Shobbrook, ~R. R.,  et al.: \mnras\ \textbf{360}, 619  (2005)

\bibitem[\protect\citeauthoryear{{Kurucz}}{2004}]{}
Kurucz, ~R.: http:// kurucz.harvard.edu (2004)

\bibitem[\protect\citeauthoryear{{Pamyatnykh et al}}{2004}]{}
Pamyatnykh, ~A. A., Handler, ~G., Dziembowski, ~W. A.: \mnras\ \textbf{350}, 1022 (2004)

\bibitem[\protect\citeauthoryear{{Sschnerr et al}}{2006}]{}
Schnerr, R. S., Verdugo, E., Henrichs, H. F., Neiner, C. 2006, \aap, \textbf{452}, 969

\bibitem[\protect\citeauthoryear{{Suarez et al}}{2009}]{}
Suarez J. C., Moya A. Amado P. J., Martin-Ruiz S.,
Rodriguez-Lopez C., Garrido R., 2009, \apj, \textbf{690}, 1401

\end{thebibliography}
\end{document}